\documentclass[journal]{IEEEtran}
\usepackage[T1]{fontenc}
\usepackage{cite}
\usepackage[dvips]{graphicx}
\usepackage{subfigure}
\usepackage{amsmath}
\DeclareGraphicsExtensions{.eps}
\usepackage{multirow}
\usepackage{color}

\begin{document}

\title{The Derived Equivalent Circuit Model for Magnetized Anisotropic Graphene}

\author{Ying~S.~Cao,~\IEEEmembership{Student Member,~IEEE,}
        Li~Jun~Jiang,~\IEEEmembership{Senior Member,~IEEE,}
        and~Albert~E.~Ruehli,~\IEEEmembership{Life~Fellow,~IEEE}

\thanks{Ying~S.~Cao and Li~Jun~Jiang are with the Department
	of Electrical and Electronic Engineering, The University of Hong Kong, Hong Kong
	(e-mail: caoying@eee.hku.hk; jianglj@hku.hk)}
\thanks{Albert~E.~Ruehli is with UMRI/MST EMC Laboratory, Missouri
  University of Science and Technology, Rolla, MO 65409, USA (e-mail: albert.ruehli@gmail.com)}}
\markboth{IEEE TRANSACTIONS ON ANTENNAS AND PROPAGATION,~Vol.~xx, No.~xx, xx~2014}
{Shell \MakeLowercase{\textit{et al.}}: Bare Demo of IEEEtran.cls for Journals}
\maketitle

\begin{abstract}
 Due to the static magnetic field, the conductivity for graphene becomes a dispersive and anisotropic tensor, which complicates most modeling methodologies. In this paper, a novel equivalent circuit model is proposed for graphene with the magnetostatic bias based on the electric field integral equation (EFIE).  To characterize the anisotropic property of the biased graphene, the resistive part of the unit circuit is replaced by a resistor in series with current control voltage sources (CCVSs).  The CCVSs account for the off-diagonal parts of the surface conductivity tensor for the magnetized graphene.  Furthermore, the definitions of the absorption cross section ($\sigma_{abs}$) and the scattering cross section ($\sigma_{sca}$) are revisited to make them feasible for derived circuit analysis.  This proposed method is benchmarked with several numerical
examples.  This paper also provides a new equivalent circuit model to deal with
dispersive and anisotropic materials.
\end{abstract}


\IEEEpeerreviewmaketitle

\section{Introduction}

\IEEEPARstart{G}{raphene}, a two-dimensional version of graphite, is a very promising material in emerging nanoelectric devices, such as transistors~\cite{transistor}, tunable terahertz (THz) antennas~\cite{THzAnte} and surface plasmon waveguides~\cite{SPW}. The surface conductivity of atomic-thick graphene denoted as $\sigma_g(\omega,~\mu_c,~t,T)$ plays a pivotal role in the surface plasmon polariton (SPP), nano-patch antenna and so on. $\sigma_g$ is a function of temperature $T$, chemical potential $\mu_c$ (dependent on carrier density, electrostatic bias, chemical doping), and relaxation time $t$. By dynamically tuning the surface conductivity, the propagation, polarization, radiation and scattering  properties of electromagnetic waves through graphene can be manipulated.

In order to model its gyrotropic effect, a two-dimensional graphene can be numerically characterized through the scalar conductivity~\cite{PEEC2015} without the magnetic bias. Some numerical methods are emerging to deal with the scalar conductivity of graphene, such as, the method of moments (MOM)~\cite{MOM}, finite difference time domain (FDTD)~\cite{FDTD11} method, and the partial element equivalent circuit (PEEC) method~\cite{Ying_AP}. For FDTD method, it turns the surface conductivity of graphene into the complex permittivity by dividing the thickness of graphene. Afterwards, by applying three dimensional meshes, the finite difference method can solve this two-dimensional geometry cumbersomely. For the integral-based algorithms, for example, the MOM method, it can directly implement the surface conductivity into the numerical process, but the physical process of how the magnetostatic field affects the properties of graphene is not clear since this is a pure mathematical process.  Among these numerical methods, PEEC has its natural advantage over any other methods. It not only proposes an equivalent circuit model for graphene which can capture the physical characteristics, but also it applies the surface conductivity graphene directly without involving volumetric meshes, which significantly reduces the time consumption and memory size.

However, with the bias by a static magnetic field, the surface conductivity of graphene becomes an anisotropic tensor~\cite{OE, AP2008}, which complicates the problem. Hence, the numerical methods have to settle the dispersive and anisotropic properties of graphene simultaneously. In~\cite{FDTD2013}, a FDTD approach is developed by transforming the 2D Drude-model surface conductivity of graphene into a volumetric tensor, and implementing it by using the auxiliary differential equation (ADE) and the matrix exponential method. However, this method uses volumetric discretizations, which slows the computation.

In this paper, a novel circuit model based on the electric field integral equation (EFIE) is proposed to solve the dispersivity and anisotropy of magnetostatically biased graphene. In the equivalent circuit model, the diagonal elements of the  surface conductivity tensor intrinsically correspond to the resistance of each inductive branch, which is the same as the unbiased scalar conductivity of graphene~\cite{PEEC2015}. For the off-diagonal elements of the conductivity tensor, a new equivalent circuit model is developed to model the resistive characteristics by utilizing current control voltage sources (CCVSs). The CCVSs are in series with the intrinsic resistors. To demonstrate our proposed method, several numerical examples are compared with the results from other numerical methods. These CCVSs in the circuit quantitively analyze how the magnetostatic field manipulates the electromagnetic characteristics of graphene. According to our best knowledge, no literature has ever developed the equivalent circuit modeling for dispersive and anisotropic graphene structures. Beyond graphene, this paper also extends PEEC method from handling isotropic materials to anisotropic and dispersive materials.

The remaining part of this paper is organized as follows: In Section II, the theory and formulations of the anisotropic conductivity tensor, and a brief  introduction to the equivalent circuit method are addressed. In Section III, details of the new equivalent circuit method for anisotropic graphene are carefully demonstrated, including derivations of the $x$-- and $y$--directional equivalent circuit model. Section IV investigates several numerical examples to benchmark our  proposed method. Conclusions are made at the end of this paper.
\section{Theory and Formulations}
\subsection{The Surface Conductivity Tensor}
\begin{figure}[htpb]
\centering
\includegraphics[width=2.5 in]{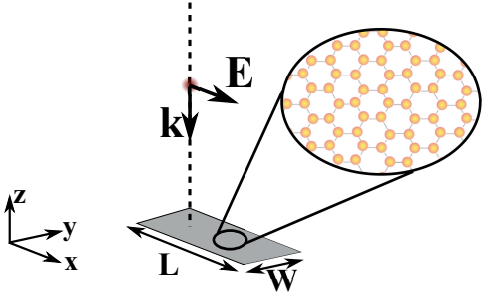}
\caption{Sketch of a graphene film placed in air, and a plane wave is linearly polarized along the length of the patch. $L$ and $W$ are the length and width of the graphene patch, respectively.}
\label{fig:graphene}
\end{figure}

 A graphene patch is placed in air, and a plane wave is linearly polarized along the length of the patch, which is illustrated in Fig.~\ref{fig:graphene}.

In the presence of the static magnetic field, the surface conductivity of graphene becomes a tensor $\overline{\overline \sigma }_g$~\cite{EMC2012, AP}. It satisfies ${\sigma _{{\rm{xx}}}} = {\sigma _{{\rm{yy}}}} = {\sigma _d}$, $-{\sigma _{{\rm{xy}}}} = {\sigma _{{\rm{yx}}}} = {\sigma _o}$. Based on this fact, the conductivity of the magnetized graphene is written as
\begin{equation}
\begin{array}{l}
{{\bar \bar \sigma }_g} 
 = \left( {\begin{array}{*{20}{c}}
{{\sigma _d}}&{ - {\sigma _o}}&0\\
{{\sigma _o}}&{{\sigma _d}}&0\\
0&0&0
\end{array}} \right)
\end{array}
\end{equation}

According to the boundary condition on the graphene surface, we have
\begin{equation}
{\bf{\hat n}} \times ({{\bf{H}}^ + } - {{\bf{H}}^ - }) = {{\bf{J}}^{surf}} = {\overline{\overline \sigma } _g}{\bf{E}}
\end{equation}
where ${{\bf{J}}^{surf}}$ is the surface current density on the graphene patch. The surface current density becomes
\begin{subequations}
\label{eq:J_surf}
\begin{align}
&J^{surf}_x = {\sigma _d}{E_x} - {\sigma _o}{E_y}
\end{align}
\begin{align}
&J_y^{surf} = {\sigma _o}{E_x} + {\sigma _d}{E_y}
\end{align}
\end{subequations}

\begin{figure}[htbp]
\centering
\includegraphics[width=2 in]{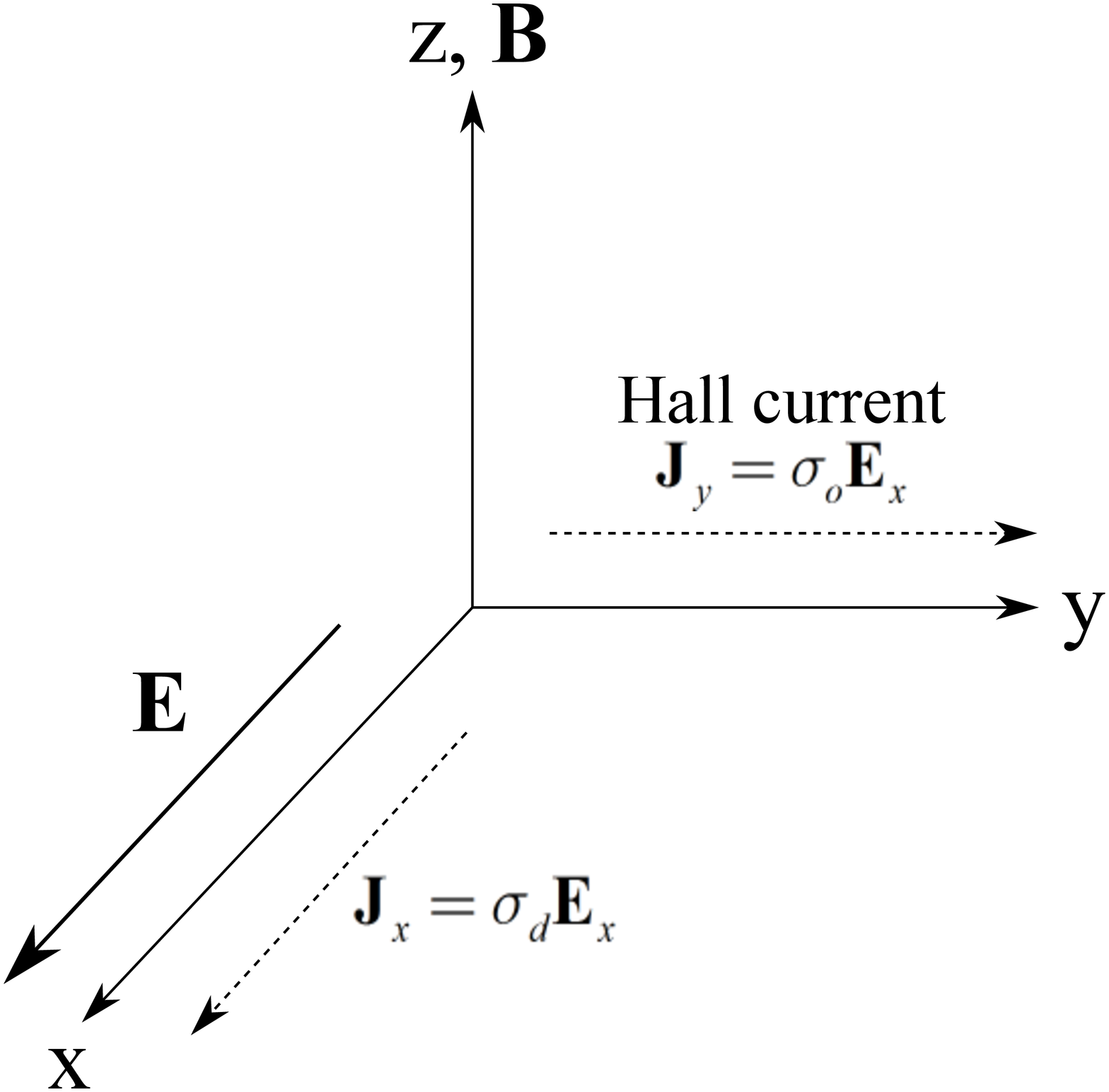}
\caption{Currents in $x$ and $y$ (according to Hall effect) directions according to electric field in $+x$--direction and anisotropic surface conductivity of graphene. The magnetic field is directed along the $+z$--axis.}
\label{fig:JxJy}
\end{figure}

In preparation for the next section, we will graphically analyze the current density $\bf{J}$ due to an electric field in the $+x$--direction, which is illustrated in Fig.~\ref{fig:JxJy}. A $y$--directional current ${\bf{J}}_y = {\sigma _o}{{\bf{E}}_x}$ is induced because of the off-diagonal term $\sigma_o$ in the conductivity tensor.

The surface conductivity is composed of both intraband and interband contributions, which is further explained in the Appendix. Even though the intraband contribution dominates the surface conductivity in the THz frequency range, the proposed method in this paper can be applied using either rigorous or  approximated formulations. In this paper, in order to improve the accuracy of the results, rigorous formulations of the conductivity (\ref{eq:exactd}) and (\ref{eq:exacto}) are applied throughout the paper.
\subsection{The Derived Circuit Based on EFIE}
According to the electric field integral equation (EFIE), the total electric field is the superposition of incident field and the scattered field. Hence, we have
\begin{equation}
\begin{split}
{{\bf{E}}^{inc}}({\bf{r}},\omega ) =& \frac{{{\bf{J}}({\bf{r}},\omega )}}{\sigma } + \mu \int_{v'} {sG({\bf{r}},{\bf{r}}'){\bf{J}}({\bf{r}}',\omega )} dv' \\
&+ \frac{\nabla }{\varepsilon }\int_{v'} {G({\bf{r}},{\bf{r}}')q({\bf{r}}',\omega )} dv'
\label{eq:efie}
\end{split}
\end{equation}
Here, $s=j\omega$, and ${G({\bf{r}},{\bf{r}}')}$ is the full-wave Green's function. The partial inductance between cell $\alpha$ and $\beta$ and partial coefficient of potential between cell $i$ and $j$ can be represented as
\begin{subequations}
\begin{align}
L{p_{\alpha \beta }} = \frac{\mu }{{{a_\alpha }{a_\beta }}}\int_{{v_\alpha }} {\int_{{v_\beta }} {G({{\bf{r}}_\alpha },{{\bf{r}}_\beta })} } d{v_\alpha }d{v_\beta }
\end{align}
\begin{align}
P{p_{ij}} = \frac{1}{{\varepsilon {S_i}{S_j}}}\int_{{S_i}} {\int_{{S_j}} {G({{\bf{r}}_i},{{\bf{r}}_j})} } d{S_i}d{S_j}
\end{align}
\end{subequations}

\begin{figure}[htbp]
\centering
\includegraphics[width=2 in]{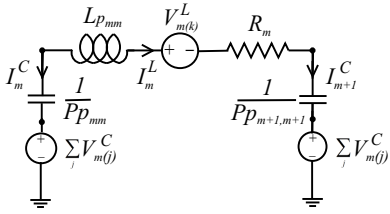}
\caption{One cell model for the $m$--th element, in which $R_m$, $Lp_{mm}$, $Pp_{mm}$ and $Pp_{m+1,m+1}$ are self resistance, inductance and self coefficients of potential for the circuit model, respectively. They correspond to $\frac{{\bf{J}}}{\sigma }$, $\frac{{\partial {\bf{A}} }}{{\partial t}}$ and $\nabla \phi$ in (\ref{eq:efie}), respectively. $V_{m(i)}^C$ and $V_{m(j)}^C$ are the voltage control voltage sources (VCVSs) due to mutual capacitive coupling between different capacitors. $V_{m(k)}^L$ is the VCVS due to mutual coupling between $Lp_{mm}$ and self inductors in other cells.}
\label{fig:PEEC_cell}
\end{figure}

Based on the electric field integral equation (EFIE), and a typical circuit model for a $m$-th cell is depicted in Fig.~\ref{fig:PEEC_cell}~\cite{Ruehli1974, Ruehli1972, Ruehli1973}. In this figure, $R_m$, $Lp_{mm}$, $Pp_{mm}$ and $Pp_{m+1,m+1}$ are self resistance, inductance and self coefficients of potential for the circuit model, respectively. They correspond to $\frac{{\bf{J}}}{\sigma }$, $\frac{{\partial {\bf{A}} }}{{\partial t}}$ and $\nabla \phi$ in (\ref{eq:efie}), respectively. $V_{m(i)}^C$ and $V_{m(j)}^C$ are the voltage control voltage sources (VCVSs) due to mutual capacitive coupling between different capacitors. $V_{m(k)}^L$ is the VCVS due to mutual coupling between $Lp_{mm}$ and self inductors in other cells. 

\section{An Equivalent Circuit Model for Anisotropic Conductivity Media}
A surface discretization is applied to represent current and charge distributions on an infinitely thin graphene sheet. By considering the tensor effect of the surface conductivity, a new circuit model can be derived for the magnetized graphene.

\begin{figure}[htbp]
\centering
\includegraphics[width=2.5 in]{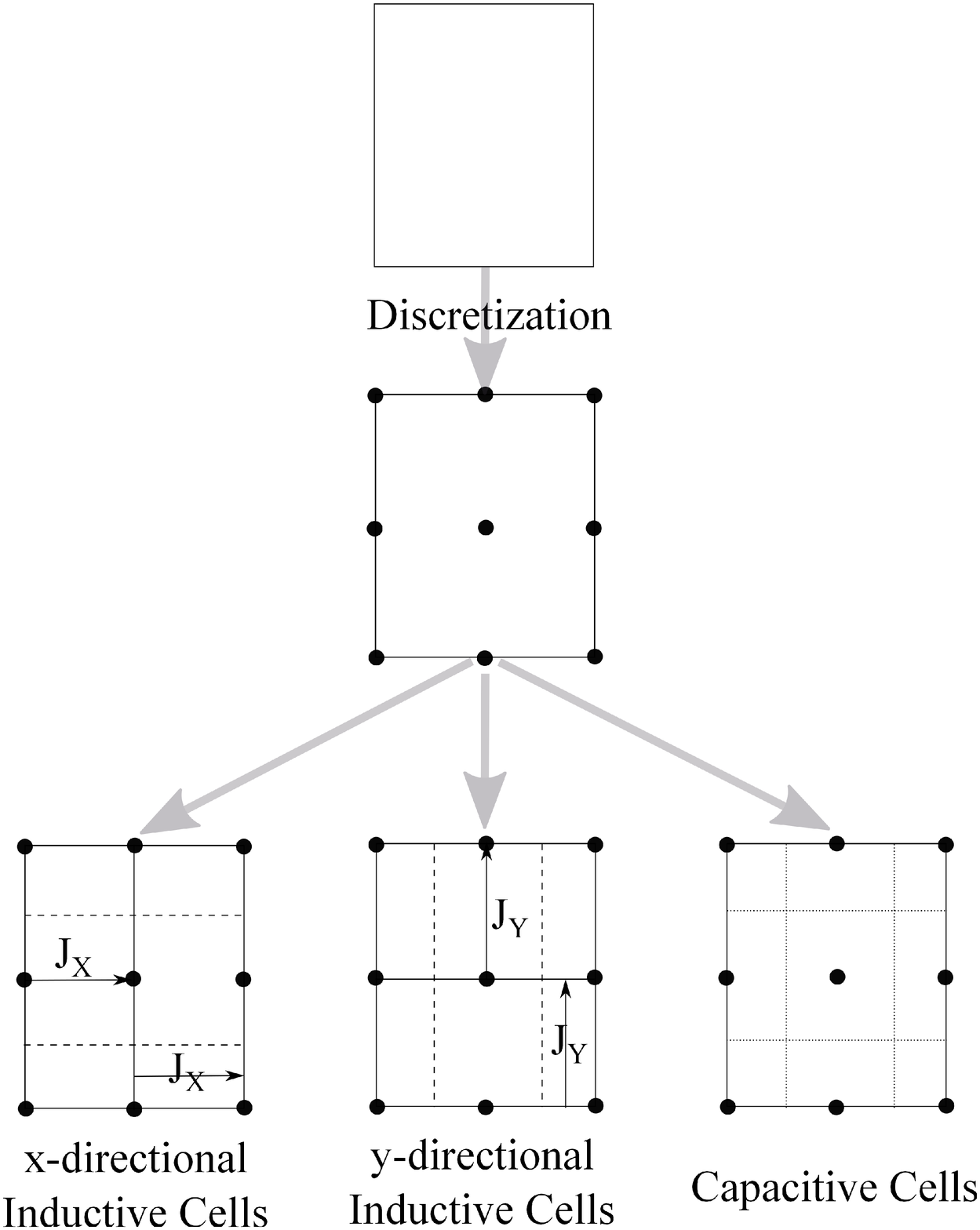}
\caption{2D discretization of thin conductive plate. Dark circles indicate nodes, dashed
lines separate inductive cells, and dotted lines separate capacitive cells.}
\label{fig:2DPEEC}
\end{figure}

Assume an infinitely thin graphene sheet is of the rectangular shape.  Its surface discretization is shown in Fig.~\ref{fig:2DPEEC}~\cite{PEEC}.  The currents ${\bf{J}}_X$ and ${\bf{J}}_Y$, are currents flowing on the graphene surface while the third current component ${\bf{J}}_Z$ is zero.

According to the superposition of electric field on a conducting plate,
\begin{equation}
\label{eq:Etot}
\begin{split}
{{\bf{E}}^{tot}} &= {{\bf{E}}^{inc}} - \frac{{\partial {\bf{A}}}}{{\partial t}} - \nabla \phi \\
 &= \overline{\overline \sigma } _g^{ - 1}{\bf{J}}\\
 &= \overline{\overline \rho } {\bf{J}}
\end{split}
\end{equation}
where $\bf{A}$ and $\phi$ are vector and scalar potentials, respectively. $\overline{\overline \rho }$ is the inverse of conductivity tensor $\overline{\overline \sigma } _g$ and denotes resistivity of graphene. Hence the total electric field is represented by $x$ and $y$ components as follows
\begin{subequations}
\label{eq:ExEy}
\begin{align}
{E_x^{tot}} = {\rho _{xx}}{J_x} + {\rho _{xy}}{J_y}
\end{align}
\begin{align}
{E_y^{tot}} = {\rho _{yx}}{J_x} + {\rho _{yy}}{J_y}
\end{align}
\end{subequations}

Combing Eq.~(\ref{eq:Etot}) and Eq.~(\ref{eq:ExEy}), we have the starting equations for circuit model derivations.
\begin{subequations}
\label{eq:JxJy}
\begin{align}
{\rho _{xx}}{J_x} + {\rho _{xy}}{J_y} + {\frac{{\partial {A_x}}}{{\partial t}} + \frac{{\partial {\phi}}}{{\partial x}} } =  - E_x^{inc}
\end{align}
\begin{align}
{\rho _{yx}}{J_x} + {\rho _{yy}}{J_y} + {\frac{{\partial {A_y}}}{{\partial t}} + \frac{{\partial {\phi}}}{{\partial y}} } =  - E_y^{inc}
\end{align}
\end{subequations}

\subsection{$x$--Directional Resistive Cell}
According to (\ref{eq:JxJy}a), the resistive contribution ${\rho _{yx}}{J_x}$ in isotropic media is changed to ${\rho _{yx}}{J_x} + {\rho _{yy}}{J_y}$. It means that the resistive effect in one direction comes from contributions of two directions. Unlike isotropic materials, magnetically biased graphene couples the currents of orthogonal directions in the lossy term. In order to get the resistive components of graphene, we need to check the two terms carefully.

\begin{figure}[htbp]
\centering
\includegraphics[width=3 in]{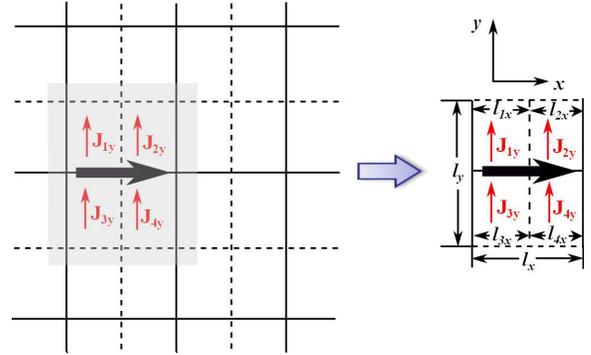}
\caption{$x$--directional graphene cell is studies first. The big black arrow shows $x$--directional current and red small arrows show $y$--directional currents. $l_x$ and $l_y$ denotes the length and width of the inductive cell, respectively. The studied inductive cell (in shadows) has constant $x$--directional current density $J_x$ and this inductive cell is distributed into four nearby $y$--directional inductive cells whose current densities are $J_{1y}$, $J_{2y}$, $J_{3y}$ and $J_{4y}$, respectively. $l_{1x}$, $l_{2x}$, $l_{3x}$ and $l_{4x}$ are the length of these current densities $J_{1y}$, $J_{2y}$, $J_{3y}$ and $J_{4y}$, respectively.}
\label{fig:PEECJx}
\end{figure}

For a surface cell, the surface current densities in $x$ and $y$ directions are defined as  $J_x=I_x/l_y$, $J_{iy}=I_{iy}/l_{ix}$, $i=1,~2,~3$ and $4$. $I_x$ is the current in $x$--directional inductive cell, and $I_{iy}$ is the current of the $i$--th $y$--directional inductive cell. $l_{ix}$ is the length of the $i$--th y-directional inductive cell. The areas of four small cells are all $l_x/2 \times l_y/2$ in Fig.~\ref{fig:PEECJx}. Hence, the voltage drop caused by the resistive loss along the $x$--directional inductive cell is defined as

\begin{equation}
\label{eq:Rx}
\begin{split}
V_R^x 
 =& {R_x}{I_x} + {R_{1x}}{I_{1y}} + {R_{2x}}{I_{2y}} + {R_{3x}}{I_{3y}} + {R_{4x}}{I_{4y}}
\end{split}
\end{equation}
where
\begin{subequations}
\begin{align}
R_x={\rho _{xx}} \frac{l_x}{l_y},
\end{align}
\begin{align}
{R_{ix}} = \frac{{{\rho _{xy}}}}{{{l_{ix}}{l_y}}}\frac{{{l_x}}}{2}\frac{{{l_y}}}{2},~i=1,~2~,~3,~\rm{and}~4.
\end{align}
\begin{align}
I_{iy}=J_{iy} l_{ix},~i=1,~2~,~3,~\rm{and}~4.
\end{align}
\end{subequations}
$R_{1x}$ -- $R_{4x}$ represent the coupled resistive effect from the $y$--direction current, and they are equivalent to the the coefficients of $x$--directional current control voltage sources (CCVSs).

\begin{figure}[htbp]
\centering
\includegraphics[width=2.5 in]{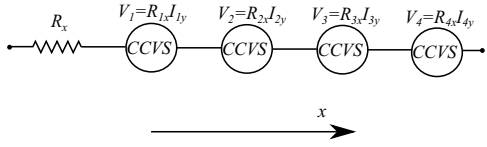}
\caption{$x$--directional resistive part of equivalent circuit model. The CCVSs are in series with a resistor.}
\label{fig:CCVSx}
\end{figure}

According to (\ref{eq:Rx}), the resistive part of the equivalent circuit model can be illustrated as a resistor in series with four CCVSs (current control voltage sources), and is shown in Fig.~\ref{fig:CCVSx}. The remaining inductive and capacitive branches are the same with that of the non-magnetized graphene~\cite{PEEC2015}.
\subsection{$y$--Directional Resistive Cell}
\begin{figure}[htbp]
\centering
\includegraphics[width=3 in]{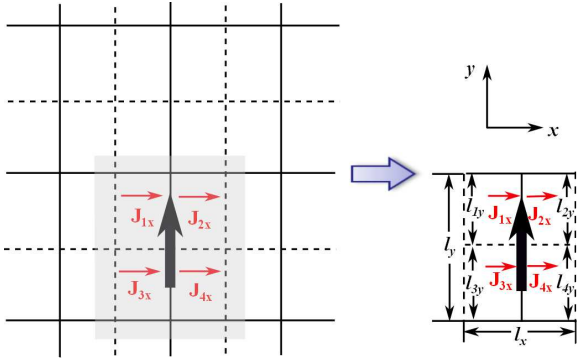}
\caption{$y$--directional graphene cell is studies first. The big black arrow shows $y$--directional current and red small arrows show $x$--directional currents. $l_x$ and $l_y$ denotes the length and width of the inductive cell, respectively. The studied inductive cell (in shadows) has constant $y$--directional current density $J_y$ and this inductive cell is distributed into four nearby $x$--directional inductive cells whose current densities are $J_{1x}$, $J_{2x}$, $J_{3x}$ and $J_{4x}$, respectively. $l_{1y}$, $l_{2y}$, $l_{3y}$ and $l_{4y}$ are the width of these current densities $J_{1x}$, $J_{2x}$, $J_{3x}$ and $J_{4x}$, respectively.}
\label{fig:PEECJy}
\end{figure}

Similarly, $y$--directional circuit model for graphene can be derived using the mesh cell shown in Fig.~\ref{fig:PEECJy}. The currents are first defined as $J_y=I_y/l_x$, $J_{ix}=I_{ix}/l_{iy}$, $i=1,~2,~3$ and $4$. $I_y$ is the current in $y$--directional inductive cell, and $I_{ix}$ is the current of the $i$--th $x$--directional inductive cell. $l_{iy}$ is the length of the $i$--th $x$--directional inductive cell. The areas of four small cells are all $l_x/2 \times l_y/2$ in Fig.~\ref{fig:PEECJy}. Hence, the voltage drop caused by the resistive loss along the y-directional inductive cell is defined as
\begin{equation}
\label{eq:Ry}
\begin{split}
V_R^y
 =& {R_y}{I_y} + {R_{1y}}{I_{1x}} + {R_{2y}}{I_{2x}} + {R_{3y}}{I_{3x}} + {R_{4y}}{I_{4x}}
\end{split}
\end{equation}
where
\begin{subequations}
\begin{align}
R_y={\rho _{yy}} \frac{l_y}{l_x},
\end{align}
\begin{align}
{R_{iy}} = \frac{{{\rho _{yx}}}}{{{l_{iy}}{l_x}}}\frac{{{l_x}}}{2}\frac{{{l_y}}}{2},~i=1,~2~,~3,~\rm{and}~4.
\end{align}
\begin{align}
I_{ix}=J_{ix} l_{iy},~i=1,~2~,~3,~\rm{and}~4.
\end{align}
\end{subequations}
$R_{1y}$ -- $R_{4y}$ represent the coupled resistive effect from the y direction current, and they are equivalent to the coefficients of
$y$--directional CCVSs.

\begin{figure}[htbp]
\centering
\includegraphics[width=0.65 in]{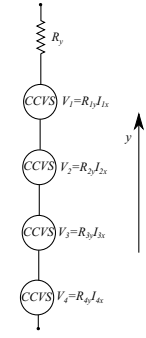}
\caption{$y$--directional resistive part of equivalent circuit model. The CCVSs are in series with a resistor.}
\label{fig:CCVSy}
\end{figure}

According to (\ref{eq:Ry}), the resistive part of the equivalent circuit model can be illustrated as a resistor in series with four CCVSs (current control voltage sources), as shown in Fig.~\ref{fig:CCVSy}. The remaining inductive and capacitive branches are the same with that of the non-magnetized graphene~\cite{PEEC2015}.

\subsection{The Complete Equivalent Circuit for Biased Graphene}
\begin{figure}[htbp]
\centering
\includegraphics[width=3.5 in]{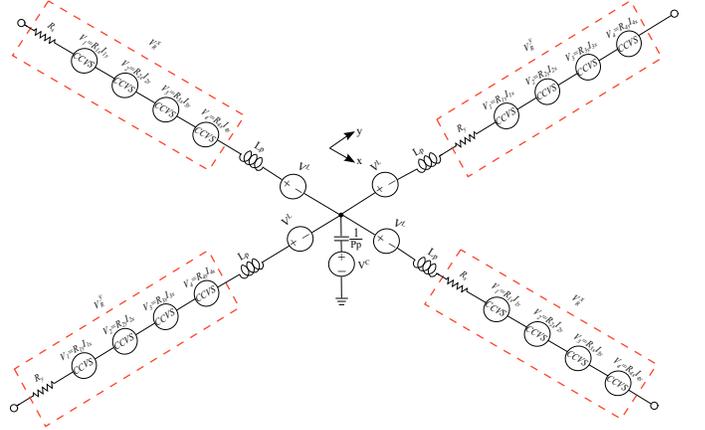}
\caption{A schematic diagram of the complete equivalent circuit for anisotropic conductivity graphene. This circuit model is for four nearby cells which share a common node. Two branches are for $x$--directional cells and the other two are for $y$--directional cells. The scripts for each cell are omitted for simplicity.}
\label{fig:unitxy}
\end{figure}

By combing the novel circuit model of $x$-- and $y$--directional resistive part of resistance, the complete equivalent circuit suitable for anisotropic conductivity media is obtained. If four unit cells are placed along the $x$ and $y$ coordinates, the equivalent circuit for these nearby cells (two in $x$-direction and two in $y$--direction) is illustrated in Fig.~\ref{fig:unitxy}.

In Fig.~\ref{fig:unitxy}, an equivalent circuit model for anisotropic graphene is sketched. The circuit elements inside the red rectangular dashed lines form the resistive part of $x$ and $y$--directional voltage drop, $V_R^x$ and $V_R^y$. This circuit model is for four nearby cells which share a common node, and two for $x$--directional cells and the other two for $y$--directional cells. In this figure, the scripts for each cell are omitted for simplicity, and all the elements in the model can be calculated according to the geometry and characteristics of each cell. This model is a combination of Fig.~\ref{fig:CCVSx} and Fig.~\ref{fig:CCVSy}.

\section{Numerical Examples}
\subsection{A Finite Size Graphene Patch}
\begin{figure}[htbp]
	\centering
	\subfigure[]{	
		\includegraphics[width=3 in]{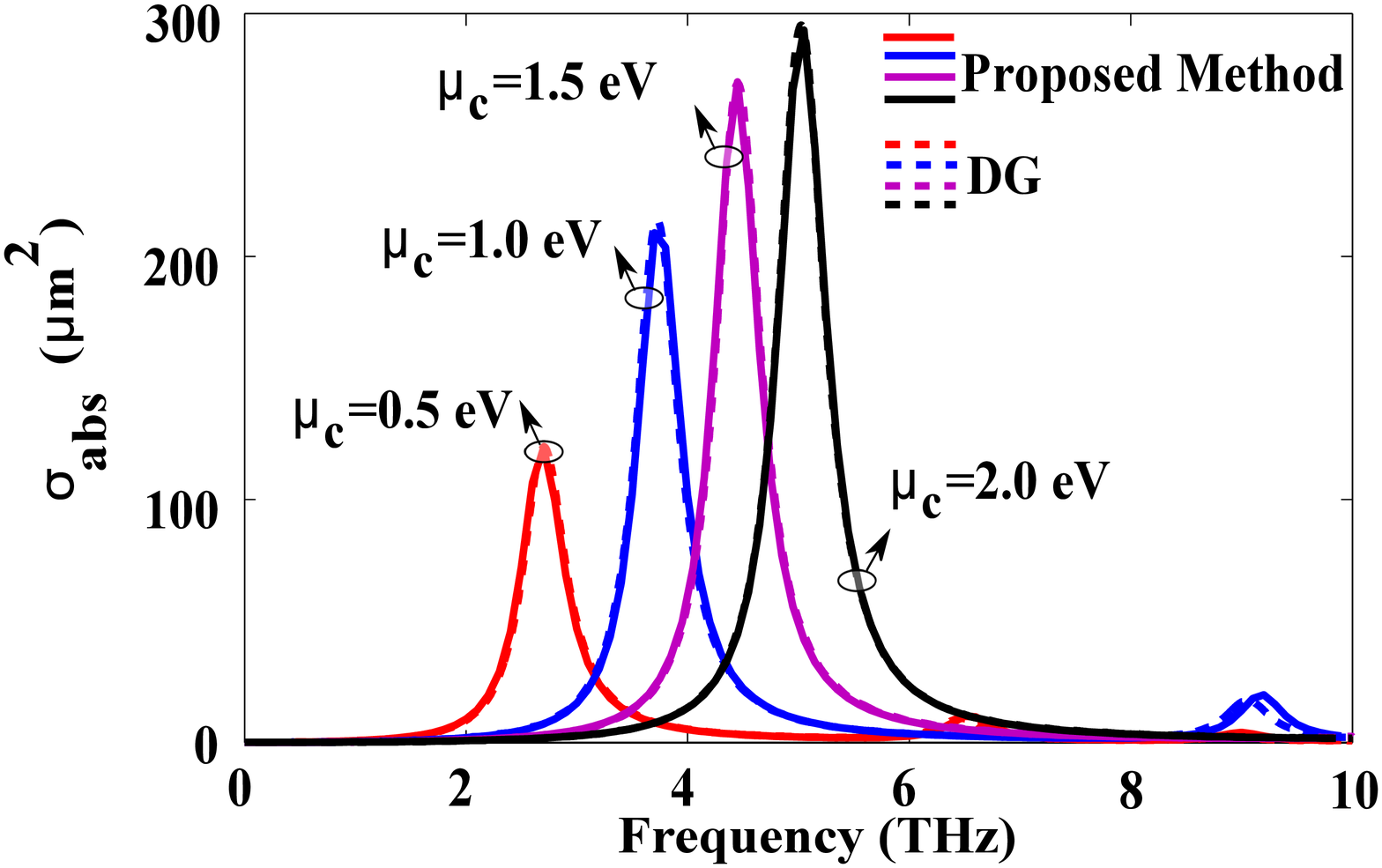}
		\label{fig:subfig1}		
	}		
	\subfigure[]{	
		\includegraphics[width=3 in]{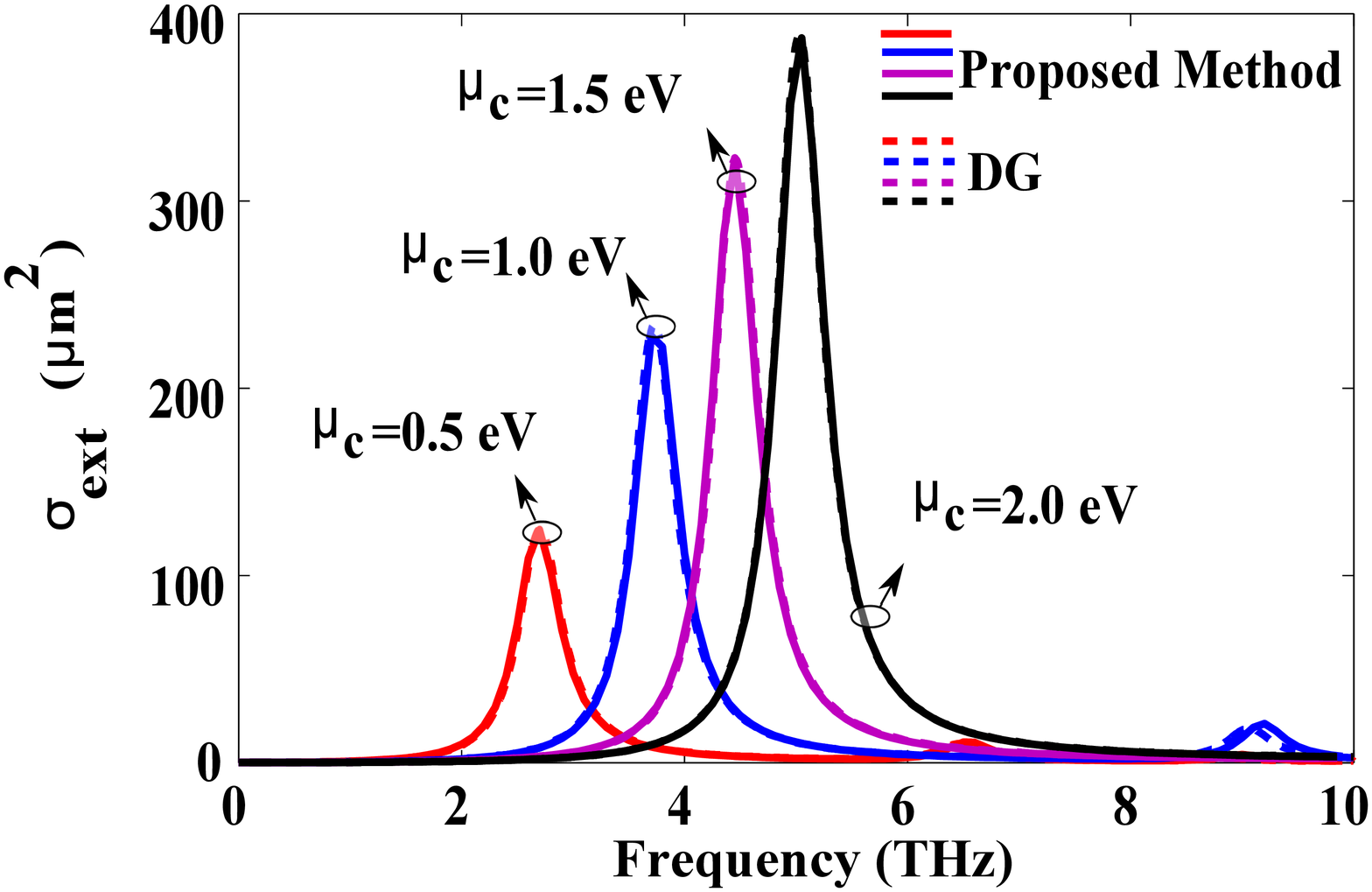}
		\label{fig:subfig2}		
	}
	\setlength{\abovecaptionskip}{0 pt}
	\caption{The comparison of (a) $\sigma_{abs}$ (absorption cross section), (b) $\sigma_{ext}$ (extinction cross section) of the graphene patch calculated with the proposed equivalent circuit method and the discontinuous Galerkin method (DG).}
	\label{fig:ACSEXT_patch}	
\end{figure}

To validate the accuracy of the proposed algorithm for the magnetized graphene, a 10 by 2 $\mu m^2$ graphene patch is studied first. The graphene patch is biased by a $z$--directional static magnetic field and is illuminated by a plane wave linearly polarized along the patch length, where ${\rm{\hat k = \hat z}}$ is the direction of propagation. Relaxation time $t=1.3 \times 10^{-13}$ s, the magnetostatic bias $B_0=0.25$ T. The absorption cross section and extinction cross section calculated by this proposed equivalent circuit model and DG~\cite{DG} (discontinuous Galerkin method) are compared in Fig.~\ref{fig:ACSEXT_patch}.

From Fig.~\ref{fig:ACSEXT_patch}, it can be concluded that the absorption cross section and extinction cross section for the magnetized graphene patch calculated by this method match very well with the discontinuous Galerkin method, which validates the proposed method in this paper.

\begin{figure}[htbp]
	\centering
	\subfigure[]{	
		\includegraphics[width=3 in]{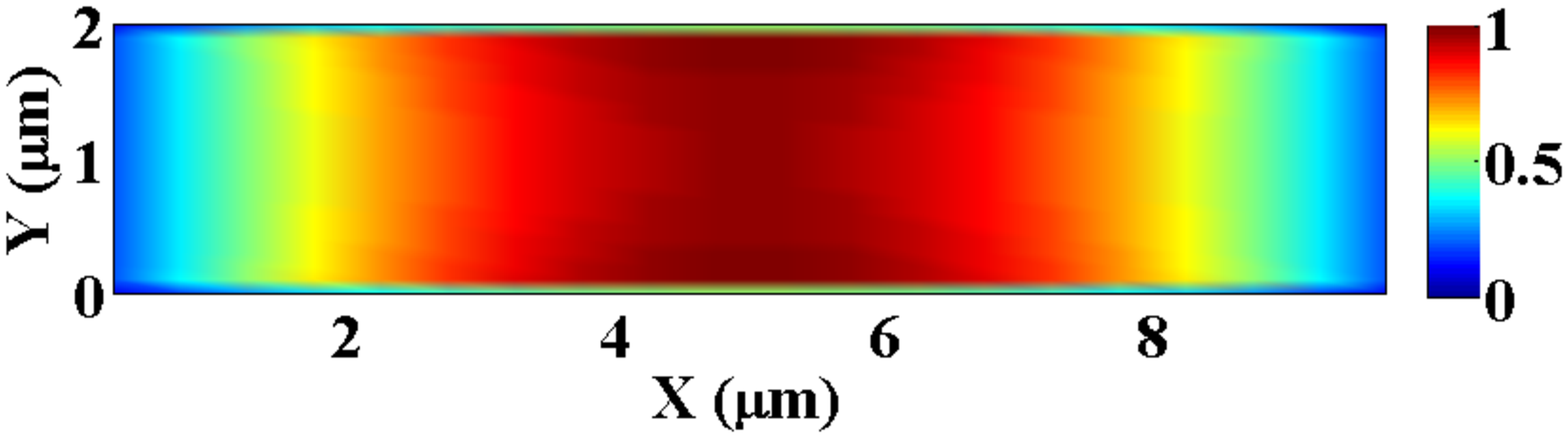}
		\label{fig:subfig1}		
	}		
	\subfigure[]{	
		\includegraphics[width=3 in]{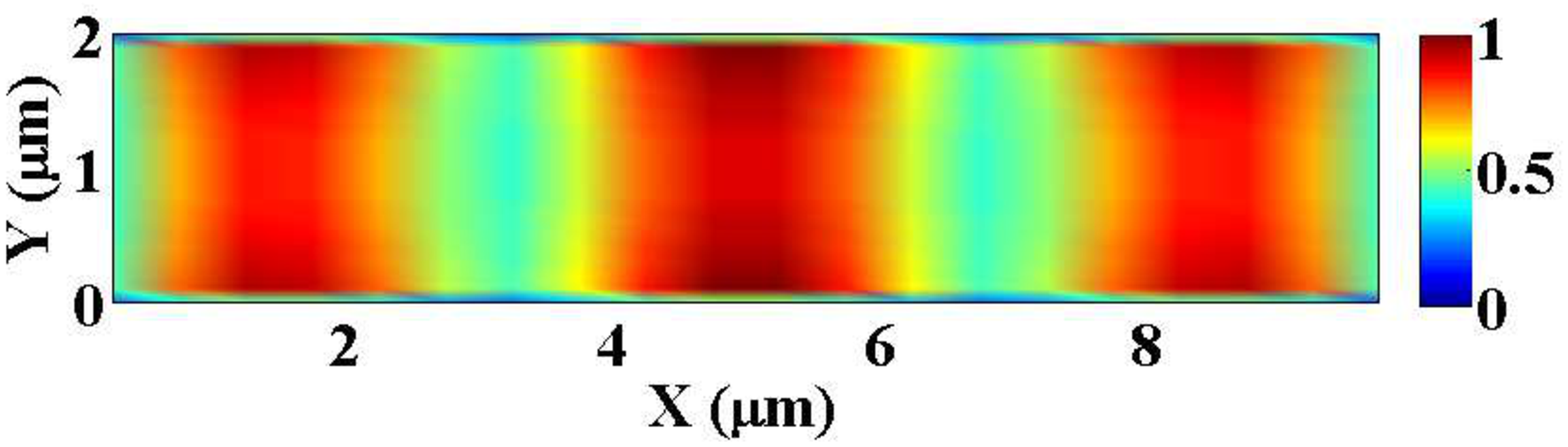}
		\label{fig:subfig2}		
	}
	\setlength{\abovecaptionskip}{0 pt}
	\caption{The comparison of normalized $x$-directional current distributions at the two resonant frequencies (a) $f_1$=3.73 THz,  (b) $f_2$=9.20 THz.}
	\label{fig:Patch_Ix}	
\end{figure}

The normalized $x$-directional current distributions on the patch are also depicted in Fig.~\ref{fig:Patch_Ix} for $\mu_c$=1.0 eV at the two resonant frequencies $f_1$=3.73 THz and $f_2$=9.20 THz. Since at the resonant frequencies, the current distributions behave like the near field electric field distribution, which illustrates that this model capture the physics very well.

\section{Conclusion}
In this paper, a novel circuit method is proposed to characterize the statically magnetized biased graphene. In order to better analyze its anisotropic and dispersive properties, a new equivalent circuit with CCVSs is developed to incorporate the off-diagonal elements of the conductivity tensor. Several examples are presented to verify this method and to illustrate the physical meanings of the CCVSs. This is the first equivalent circuit-based approach to deal with anisotropic graphene according to the authors' knowledge.
\section*{Acknowledgment}
This work was supported in part by the Research Grants Council of Hong Kong (GRF 712612 and 711511), NSFC 61271158, HKU Seed Fund 201309160052, and Hong Kong UGC AoE$/$P--04$/$08.
The authors are also grateful for the helpful comments from
reviewers, editors, and for the constructive suggestions from
Prof. W. C. Chew.

\appendices
\section{the Conductivity Tensor}
\subsection{Rigorous Formula of the Surface Conductivity Tensor}
By counting both intraband and interband contributions of graphene, the rigorous explicit expressions of ${\sigma _d}$ and ${\sigma _o}$ in (\ref{eq:J_surf}) are~\cite{magneto}
\begin{equation}
\label{eq:exactd}
\begin{split}
&{\sigma _d}({\mu _c}({E_0}), {B_0})\\
  =&  \frac{{{e^2}v_{\rm{F}}^2\left| {e{B_0}}  \right|(\omega  - j2\Gamma )\hbar }}{{ - j\pi }}\\
  &\times \sum\limits_{n = 0}^\infty   {\{ \frac{{{f_d}({M_n}) - {f_d}({M_{n + 1}}) + {f_d}( - {M_{n + 1}}) - {f_d}( - {M_n})}}{{{{({M_{n + 1}} - {M_n})}^2} - {{(\omega  - j2\Gamma )}^2}{\hbar ^2}}}} \\
 & \times (1 - \frac{{{\Delta ^2}}}{{{M_n}{M_{n + 1}}}})\frac{1}{{{M_{n + 1}} - {M_n}}}\\
 & + \frac{{{f_d}( - {M_n}) - {f_d}({M_{n + 1}}) + {f_d}( - {M_{n + 1}}) - {f_d}( - {M_n})}}{{{{({M_{n + 1}} - {M_n})}^2} - {{(\omega  - j2\Gamma )}^2}{\hbar ^2}}}\\
 & \times (1 + \frac{{{\Delta ^2}}}{{{M_n}{M_{n + 1}}}})\frac{1}{{{M_{n + 1}} + {M_n}}}\}
\end{split}
\end{equation}

and
\begin{equation}
\label{eq:exacto}
\begin{split}
& {\sigma _o}({\mu _c}({E_0}),{B_0})\\
  =&  - \frac{{{e^2}v_{\rm{F}}^2\left| {e{B_0}} \right|}}{\pi }\\
 & \sum\limits_{n = 0}^\infty   {\{ {f_d}({M_n}) - {f_d}({M_{n + 1}}) + {f_d}( - {M_{n + 1}}) - {f_d}( - {M_n})} \} \\
  &\times  \{ (1 - \frac{{{\Delta ^2}}}{{{M_n}{M_{n + 1}}}})\frac{1}{{{{({M_{n + 1}} - {M_n})}^2} - {{(\omega  - j2\Gamma )}^2}{\hbar ^2}}}\\
 &\times  (1 + \frac{{{\Delta ^2}}}{{{M_n}{M_{n + 1}}}})\frac{1}{{{{({M_{n + 1}} - {M_n})}^2} - {{(\omega  - j2\Gamma )}^2}{\hbar ^2}}}\}
\end{split}
\end{equation}
where
\begin{equation}
{M_n} = \sqrt {{\Delta ^2} + 2nv_F^2\left| {e{B_0}} \right|\hbar }
\end{equation}
and $-e$ is the charge of an electron, $\hbar$ is the reduced Planck's constant, $B_0$ is the static magnetic field, and $f_d$ is the Fermi-Dirac distribution. And
\begin{equation}
{f_d}(\varepsilon ) = {({e^{(\varepsilon  - {\mu _c})/{k_B}T}} + 1)^{ - 1}}
\end{equation}
where $\varepsilon$ is energy, and $k_B$ is the Boltzmann's constant.
\subsection{Intraband Approximation}
The analytical expressions for surface conductivity $\sigma_d$ and $\sigma_o$ are composed of intraband and interband contributions. The rigorous explicit expressions of ${\sigma _d}$ and ${\sigma _o}$ in (\ref{eq:J_surf}) are~\cite{magneto}. For frequencies within THz band, the intraband term is dominant in the total conductivity. For simplicity, the expressions of $\sigma_d$ and $\sigma_o$ can be approximated by a Drude-like model~\cite{gyrotropy, FDTD11}, which are
\begin{subequations}
\label{eq:intra}
\begin{align}
 {\sigma _d} = \sigma \frac{{1 + j\omega t}}{{{{({\omega _c}t)}^2} + {{(1 + j\omega t)}^2}}}
\end{align}
\begin{align}
{\sigma _o} = \sigma \frac{{{\omega _c}t}}{{{{({\omega _c}t)}^2} + {{(1 + j\omega t)}^2}}}
\end{align}
\end{subequations}
with
\begin{equation}
\sigma  = \frac{{{e^2}t{k_B}T}}{{\pi {\hbar ^2}}}\left[ {\frac{{{\mu _c}}}{{{k_B}T}} + 2\ln ({e^{ - {\mu _c}/{k_B}T}} + 1)} \right]
\end{equation}
where $T$ is the Kelvin temperature, $t$ is the relaxation time, and $\omega_c$ is the cyclotron frequency with ${\omega _c} = e{B_0}v_{\rm{F}}^2/{\mu _c}$ and $v_F=10^6$ m/s denotes the Fermi velocity.

\section{Extinction, Absorption and Scattering Cross Sections}
The extinction power is the total power removed from the incident field (the sum of the absorbed and the scattered powers) due to the presence of the scattering object $\Omega$. The absorbed power is the power flowing into the body and the power scattered from an object $\Omega$ is given by the real part of the integral of the outward-directed normal component of ${{\bf{S}}_{sca}}$ over ${\partial \Omega }$. ${{\bf{S}}_{tot}}$ is the time average Poynting vector. ${{\bf{S}}_{inc}}$, ${{\bf{S}}_{sca}}$ and ${{\bf{S}}_{ext}}$ are incident, scattered and extinguished components of the Poynting vector. According to \cite{abssca_light}, $P_{ext}=P_{sca}+P_{abs}$.
\begin{subequations}
\begin{align}
 {P_{abs}} = \mathop{{\int\!\!\!\!\!\int}\mkern-21mu \bigcirc}\limits_{\partial \Omega }
 {\left\langle {{{\bf{S}}_{tot}}} \right\rangle  \cdot {\bf{\hat n}}}
 \end{align}
 \begin{align}
 {P_{sca}} = \mathop{{\int\!\!\!\!\!\int}\mkern-21mu \bigcirc}\limits_{\partial \Omega }
 {\left\langle {{{\bf{S}}_{sca}}} \right\rangle  \cdot {\bf{\hat n}}}.
 \end{align}
\end{subequations}
According to the distributed power characterization~\cite{EMC}, the absorbed and scattered power can be represented by the current distributions and the partial elements, which are
\begin{subequations}
\begin{align}
P_{abs}=P^{ohm}
\label{eq:Pabs}
\end{align}
\begin{align}
P_{sca}=P^r
\label{eq:Psca}
\end{align}
\end{subequations}
where $P^r$ and $P^{ohm}$ are the radiated power and ohmic power loss.  They can be conveniently represented and calculated based on the new method in~\cite{EMC}, without involving field calculation and far field integration. For instance, the radiated and ohmic power from the segmentation $a$ is formulated as:
\begin{subequations} \label{eq:radiate}
	\begin{align}
		P_{a(a)}^{r'} =& {\rm \frac{1}{2}}{\rm{Re}}( {{\bf I}_a^L}^{\dagger} {  {\bf \overline{Z}}_L^{aa}}  {{\bf I}_a^L}  + {{\bf I}_a^C}^{\dagger} {\overline {\bf Z}_C^{aa}}  {{\bf I}_a^C} ).
	\end{align}
\begin{align}
P_a^{ohm}=&{\rm \frac{1}{2}}{{\bf I}_a^L}^{\dagger}{  {\bf \overline{Z}}_{ohm}^{aa}}  {{\bf I}_a^L}
\end{align}
\end{subequations}
where the superscript `$\dagger$' means the conjugate transpose of a
vector. ${\bf I}_a^L$  is the vector of current distribution of inductive branches on $a$.  $\overline {\bf Z}_L^{aa}$ is an $N_{L_a} \times N_{L_a}$ matrix in which all elements are inductive coupling impedances ($N_{L_a}$ and is the number of inductive cells on $a$. $\overline {\bf Z}_{L(mk)}^{aa}=j \omega Lp_{mk}$), $\overline {\bf Z}_C^{aa}$ is an $N_{C_a} \times N_{C_a}$ matrix whose elements
correspond to capacitive coupling ($N_{C_a}$ is the
number of capacitive cells on $a$. $ \overline
{\bf Z}_{C(mk)}^{aa}=\frac {Pp_{mk}} {j \omega}$).
${{\bf \overline{Z}}_{ohm}^{aa}}$ is a diagonal matrix which contains the resistive part of each cell.

Based on the power analysis~\cite{EMC}, the absorption, scattering and extinction cross section are reformulated as
\begin{subequations}
\label{eq:sigma}
\begin{align}
&{\sigma _{abs}} = \frac{{{P_{abs}}}}{{\left| {{{\bf{S}}_{inc}}} \right|}}.
\end{align}
\begin{align}
&{\sigma _{sca}} = \frac{{{P_{sca}}}}{{\left| {{{\bf{S}}_{inc}}} \right|}}.
\end{align}
\begin{align}
&{\sigma _{ext}}={\sigma _{abs}}+{\sigma _{sca}}.
\end{align}
\end{subequations}

Equation~(\ref{eq:radiate}) - (\ref{eq:sigma}) convert the calculation process for cross sections into an intrinsic circuit problem that can be solved without involving direct electric and magnetic field calculations.



%

\end{document}